\documentclass[12pt]{article}

\usepackage{epsfig}

\begin{document}

\newcommand{\beq}{\begin{equation}}
\newcommand{\eeq}{\end{equation}}
\newcommand{\beqa}{\begin{eqnarray}}
\newcommand{\eeqa}{\end{eqnarray}}

\def\ov{\overline}
\def\onlyif{\rightarrow}

\def\openone{\leavevmode\hbox{\small1\kern-3.8pt\normalsize1}}

\def\a{\alpha}
\def\b{\beta}
\def\g{\gamma}
\def\r{\rho}
\def\minus{\,-\,}
\def\eks{\bf x}
\def\kay{\bf k}

\def\ket#1{|\,#1\,\rangle}
\def\bra#1{\langle\, #1\,|}
\def\braket#1#2{\langle\, #1\,|\,#2\,\rangle}
\def\proj#1#2{\ket{#1}\bra{#2}}
\def\expect#1{\langle\, #1\, \rangle}
\def\trialexpect#1{\expect#1_{\rm trial}}
\def\ensemblexpect#1{\expect#1_{\rm ensemble}}
\def\kpsi{\ket{\psi}}
\def\kphi{\ket{\phi}}
\def\bpsi{\bra{\psi}}
\def\bphi{\bra{\phi}}

\def\ditto{\rule[0.5ex]{2cm}{.4pt}\enspace}
\def\th{\thinspace}
\def\ni{\noindent}
\def\thirty{\hbox to \hsize{\hfill\rule[5pt]{2.5cm}{0.5pt}\hfill}}

\def\set#1{\{ #1\}}
\def\setbuilder#1#2{\{ #1:\; #2\}}
\def\Prob#1{{\rm Prob}(#1)}
\def\pair#1#2{\langle #1,#2\rangle}
\def\Id{\bf 1}

\def\dee#1#2{\frac{\partial #1}{\partial #2}}
\def\deetwo#1#2{\frac{\partial\,^2 #1}{\partial #2^2}}
\def\deethree#1#2{\frac{\partial\,^3 #1}{\partial #2^3}}

\newcommand{\xx}{{\scriptstyle -}\hspace{-.5pt}x}
\newcommand{\yy}{{\scriptstyle -}\hspace{-.5pt}y}
\newcommand{\zz}{{\scriptstyle -}\hspace{-.5pt}z}
\newcommand{\kk}{{\scriptstyle -}\hspace{-.5pt}k}
\newcommand{\sx}{{\scriptscriptstyle -}\hspace{-.5pt}x}
\newcommand{\sy}{{\scriptscriptstyle -}\hspace{-.5pt}y}
\newcommand{\sz}{{\scriptscriptstyle -}\hspace{-.5pt}z}
\newcommand{\sk}{{\scriptscriptstyle -}\hspace{-.5pt}k}

\def\openone{\leavevmode\hbox{\small1\kern-3.8pt\normalsize1}}

\title{Implementing the modular eballot system V0.6}
\author{Andrea Pasquinucci
\\
\small
{\it {\rm UCCI.IT}, via Olmo 26, I-23888 Rovagnate (LC), Italy
}}
\date{December 22, 2006}
\maketitle

\abstract{We describe a practical implementation of the modular
eballot system proposed in ref.~\cite{eballot}.}

\vspace{1 cm} 
\normalsize

\section{Introduction}

Implementing a balloting system requires making many choices when the 
theoretical steps of the protocol must be realized by software 
procedures. Simple statements can become quite complex to implement and 
many subtle points arise which could reduce or make useless the entire 
protocol.

For this reason, we consider to be a very important issue to implement 
in a working system the protocol proposed in ref.\ \cite{eballot}. We 
will see as some a-priori simple steps are quite problematic to 
implement and we discuss our solutions and possible improvements.

The paper is organized as follows. In section 2 we describe the approach 
taken in building the system, in section 3 the hardware and software 
choices, in section 4 the procedure to setup the ballot, in section 5 
the setup of the operating system and in section 6 the setup of the 
application.

\section{Approach}

To implement the {\sl modular eballot} protocol we can choose basically 
one of the following approaches:

\begin{enumerate}

\item design, develop and implement custom hardware and software

\item design, develop and implement custom software based on 
commercially available hardware

\item use common hardware and software and design, develop and implement 
only the specific software applications needed to implement the 
protocol.

\end{enumerate}

If we consider these three approaches purely theoretically, they are 
obviously ordered in decreasing level of security. Indeed a custom 
hardware/software solution can in principle offer the maximum possible 
security.

But in practice the effort required by approaches 1 and 2 is often too 
large. Besides the need for a large team of developers and long 
developing times, the probability that at least for the first few 
releases/models there will be many vulnerabilities, is very high.

In our setup with limited resources and short developing times, we have 
adopted the third approach, choosing common hardware and well proved 
software to implement the eballot protocol. We have selected well known 
and well supported software, which have known history of vulnerabilities 
and fast releases of patches.

We have then chosen software that we can {\sl trust} in the sense that 
they have a very large number of users and large support, trading the 
low number of vulnerabilities and external support for the absence of 
features which could be helpful in implementing the protocol and the 
presence of features not needed for it.

\section{The hardware/software choices}

For the first release of the implementation we have chosen the simplest 
possibility:

\begin{itemize}

\item the voter uses her own web browser

\item the AuthSrv and the VoteSrv are implemented as web applications 
running on common web servers

\item the Anonymizer is a simple NAT device.

\end{itemize}

With this setup it is not possible to implement the blind signature 
version of the protocol presented in ref.\ \cite{bruschi} and discussed 
in Appendix A of ref.\ \cite{eballot}. We will not consider blind 
signatures for our implementation nor in the rest of this paper.

An improvement on the current implementation would be to realize the 
{\sl client proxy}. This will allow to add more features for the voter, 
like a more controlled access to the system, automatic check of the 
digital certificate of the web sites, removal at the origin of all 
personal information from the http transaction, like the UserAgent 
field. Notice that the realization of a client proxy will require:

\begin{itemize}

\item to support versions of the client proxy for all computing 
platforms that the voters could use

\item the possibility of vulnerabilities in the client proxy and the 
necessity to implement a system to manage its patches, distribution of 
new releases etc.

\end{itemize}

For what concerns the Anonymizer, a simple approach to improve over the 
simple NAT device is to make the voters access the web servers through 
the {\sl tor} network \cite{tor}.

In any case, the voter (or the client proxy for her) must check the 
fingerprints of the digital certificates of the AuthSrv and VoteSrv to 
prevent man-in-the-middle attacks.

For what concerns the Operating System, the choice has been dictated 
mostly by our experience and personal competence and has been a Linux 
distribution. The system can be implemented almost identically on any 
*BSD distribution or Unix-like OS. What in particular lead us to the 
choice of Linux has been the RSBAC kernel patch \cite{rsbac} which 
provides Mandatory Access Control (MAC) features very useful for the 
security of the servers and for implementing some aspects of the 
protocol. 

As for the web server application, it has obviously been chosen Apache 
\cite{apache} with PHP \cite{php}. For added simplicity, no SQL server 
is used but all data is stored in flat files. For SSL/TLS openssl 
\cite{openssl} is used whereas for the cryptographic operations of the 
protocol the gnupg \cite{gnupg} implementation of OpenPGP \cite{openpgp} 
with the gpgme and gpg-agent extensions, has been adopted.

Clocks of the servers are synchronized using ntp \cite{ntp} but the 
ntpd daemon is not running during a ballot.

During a ballot the only open ports on the servers are tcp 80/443 (and,
only if needed, icmp echo-request). 
The only possible way of login on the servers is at the console. 

The core of the implementation is given by PHP-5 scripts which can be 
downloaded under a GPLv2 license. The PHP scripts require the php-gpgme 
extension to interact with gnupg. Some crucial routines have been 
written in C. 

In the rest of this paper we will describe our example implementation of 
the protocol, even if the system can be installed on different OS and 
web-servers following a similar procedure, or implemented in similar 
way.

\section{Ballot setup and procedures}

To implement the protocol, besides the voter, 5 human roles, 
ballot officials and machine administrators, are needed:

\begin{enumerate}

\item {\sl the Authentication Managers (AuthMgr)} must organize the
ballot selecting the eligible voters and providing them with the voting
credentials; at the end of the voting period they must check that only
eligible voters have casted a ballot

\item {\sl the Managers of the Authentication Web Server (AuthSysMgr)}
must setup and run the Authentication Web Server and provide the AuthMgr
with the list of voters that have casted a ballot at the end of the
voting period

\item {\sl the Managers of the Anonymizer System (AnonSysMgr)} must
guarantee that all connections to the Vote Web Server do not leak
information on the IP address of the voter computer

\item {\sl the Vote Managers (VoteMgr)} must count the votes at the end
of the voting period and announce the result of the ballot

\item {\sl the Managers of the Vote Web Server (VoteSysMgr)} must setup 
and run the Vote Web Server and provide, at the end of the voting 
period, the VoteMgr with the encrypted votes that have been casted. In 
the configuration of the Vote Web Server the VoteSysMgr must be careful 
that no information on the voter's connection will be ever logged (this 
is an extra precaution since this information should be removed by the 
client-proxy and the Anonymizer).

\end{enumerate}

Each manager creates her own OpenPGP private/public key and gives to the 
other managers the public key. 

Every cryptographic operations must be done with reasonably secure 
algorithms, so for example symmetric algorithms, like AES, should have 
at least 128-bit keys, asymmetric algorithms, like RSA, at least 
1024-bit keys and cryptographic digest (or hashes) more than 130-bit 
hash, so at the moment of writing for example SHA1 can be used but not 
MD5.

The AuthMgr, optionally in collaboration with the AuthSysMgr, creates 
the vote credentials for all voters:

\begin{itemize}

\item {\sl username + password} and/or client digital certificate (these can 
be reused for more ballots)

\item a PseudoRandom (PR) string called {\sl VoteToken},\footnote{This is also 
called a {\sl Secret Token} to stress the role of this credential.} 
unique for each voter and each ballot and that can be used only once.

\end{itemize}

(As a technical detail, the username have been chosen in the charset
[a-zA-Z0-9\_.@-] and the VoteToken in the charset [a-zA-Z0-9\_.]~.)

The AuthSysMgr and the VoteSysMgr create the digital certificates for 
their https web servers and give the fingerprints of the certificate to 
the AuthMgr. 

The AuthMgr distributes to all voters the ballot credentials and the 
fingerprints. It is suggested that the username+password and VoteToken 
are distributed using different communication channels.

The AuthSysMgr installs in the gnupg keyring of the AuthSrv her own 
private and public OpenPGP key and the public keys of the AuthMgr and 
VoteSysMgr. The VoteSysMgr does the same on her server with her own 
private and public OpenPGP key and the public keys of the AuthMgr, 
AuthSysMgr and VoteMgr.

The AnonSysMgr must configure the Anonymizer so to redirect all packets 
to the VoteSrv and not log any connection.

After the applications have been configured (see the next sections) the 
server will be sealed by the AuthMgr and VoteMgr respectively and 
no login or modification of the software can be done. All accesses or 
modification must be logged and reported so that the respective manager 
can check the integrity of the machine during all the period of the 
ballot. After the sealing of the servers, the systems start accepting 
connections from the voters.

After a voter has successfully authenticated, the AuthSrv checks that 
the VoteToken has not been used already and generates a unique PR string 
called {\sl VoteAuthorization} (the VoteAuthorization has been chosen in 
the charset [a-zA-Z0-9\_.]). The AuthSrv digitally signs and encrypts 
with the AuthMgr public key the used VoteToken together with the 
username of the voter and the current timestamp, and saves it on the 
disk in a file having as name the VoteToken to mark it as used. The 
AuthSrv digitally signs and encrypts with the VoteSysMgr public key the 
VoteAuthorization (and the optional PIN, see \cite{eballot}) and sends 
it to the voter (optionally with the PIN in clear). The AuthSrv also 
digitally signs and encrypts with the AuthMgr public key the 
VoteAuthorization (but not the optional PIN) and saves it on the disk in 
a file having as name the VoteAuthorization.

The voter then connects to the VoteSrv through the Anonymizer, receives 
the web Form with the ballot, and sends to the server her ballot and the 
VoteAuthorization. The VoteSrv verifies that the VoteAuthorization 
decrypts correctly, has the correct signature of the AuthSysMgr, and has 
not been used before. If there is a PIN, it checks that the PIN 
submitted in the Form is the same as the one included in the decrypted 
VoteAuthorization. The VoteSrv digitally signs and encrypts with the 
AuthMgr public key the VoteAuthorization and saves it on the disk in a 
file having as name the VoteAuthorization to mark it as used. It then 
computes a digest of the vote with the current timestamp and a random string 
and calls it {\sl VerificationCode}, 
digitally signs and encrypts with the VoteMgr 
public key the vote together with the VerificationCode and saves it on 
the disk in a file having as name the VerificationCode (the 
VerificationCode is a hexadecimal string). Finally it computes the 
digital signature of the VerificationCode and sends to the voter the 
VerificationCode, its digital signature, the timestamp and the random string. 

The voter can recompute the VerificationCode using her vote, the 
timestamp and the random string, 
and the VoteMgr can check that the VerificationCode is 
authentic by verifying the digital signature of the VerificationCode 
done by the VoteSrv.

At the end of the ballot period, the AuthMgr checks the AuthSrv and 
removes the seal. Then the AuthSysMgr prepares one CD with the list of 
used VoteTokens and created VoteAuthorizations and gives it to the 
AuthMgr. The VoteMgr checks the VoteSrv and removes the seal. The 
VoteSysMgr prepares two CDs, one with the used VoteAuthorizations and 
gives it to the AuthMgr, the other with the votes and gives it to the 
VoteMgr.

The AuthMgr decrypts and checks the signatures of her two CDs. She 
verifies that the information is consistent and that the number of votes 
is consistent with the number of used VoteTokens and VoteAuthorizations. 
The AuthMgr publishes the list of used VoteTokens with the associated 
username and timestamp and the list of unused VoteTokens. 

The VoteMgr does not decrypts the votes but publishes the list of 
VerificationCodes (remember that each vote is in a file having as name 
the VerificationCode). 

The voters have now time to check that the information published are 
correct: if they have voted or not, the time of their authentication 
and the existence of their VerificationCode. 
If there is a problem they must report it to the AuthMgr or VoteMgr. 

After this check is done, the VoteMgr decrypts the votes, checks the 
signatures and count them. The VoteMgr then publishes the list of each 
vote with the associated VerificationCode and the final results of the 
ballot. Each voter can check that her vote is correctly listed and that 
the final count is correct.

We now assume that the systems are setup correctly and that the 
procedures are followed as described. 

\section{Operating system setup and security}

As already mentioned, the two servers must be configured so that they 
can be {\sl sealed} before the beginning of the ballot and the 
managers can check that no modification to the software nor login to the 
system are done during the ballot itself. We will describe below how we 
have chosen to seal the systems. For what concerns the verification of 
the integrity two things must be done:

\begin{itemize}

\item verify that programs are not modified

\item verify that logs are not tampered with.

\end{itemize}

For the integrity of the programs and of the log files various 
approaches can be adopted. For example a forensic image of the disk can 
be made before the sealing of the machine and a verification of the 
modifications after the end of the ballot. It is simpler, but we believe 
still appropriate for our purposes, to use a program like tripwire 
\cite{tripwire} or aide \cite{aide} to record the digest and properties 
of all files on the disk and check that they have not been modified at 
the end of the ballot.

Besides the normal hardening of the operating system, we have included 
some extra features which should increase notably the level of security 
of the systems.

As mentioned, for our implementation we have adopted the RSBAC patch to 
the Linux kernel which introduces Mandatory Access Control features. The 
RSBAC-patched Linux kernel, for our purposes, can be said to be in one 
of the following three states:

\begin{description}

\item[rsbac\_softmode] MAC rules are not enforced but violations are 
logged

\item[rsbac\_enforcing] MAC rules are enforced at the kernel level but 
the security officer (uid=400) can modify the MAC rules (the root user, 
uid=0, cannot modify the MAC rules, nor any other user)

\item[rsbac\_frozen] MAC rules are enforced at the kernel level and 
nobody can modify any MAC rule: to switch to one of the previous states 
a reboot is needed with a special kernel parameter (this is the default 
state).

\end{description}

\noindent The rsbac\_frozen mode corresponds to the sealed status of the 
servers. 

The MAC rules adopted mark most of the disk as read-only and execute, 
mark the log files as append-only, disable the loading of kernel modules 
after boot, disable the mounting of partitions on directories etc. In 
particular, all executable and configuration files both of the OS and of 
the eballot application (e.g.\ the PHP scripts) are marked as read-only 
and executable so that they cannot be modified while running in 
enforcing or frozen mode by any user or program.

But the specific purpose of RSBAC is to solve a particular problem of 
our implementation. As we have seen both the AuthSrv and the VoteSrv 
record information in flat files:

\begin{itemize}

\item the AuthSrv writes on disk the used VoteTokens

\item the AuthSrv writes on disk the created VoteAuthorizations

\item the VoteSrv writes on disk the used VoteAuthorizations

\item the VoteSrv writes on disk the votes.

\end{itemize}

\noindent Each one of these is written to its own file with its own 
distinguished name. We should guarantee that

\begin{itemize}

\item the file is written only if it does not exist

\item the file cannot be modified or deleted after creation.

\end{itemize}

\noindent In practice we need write-once read-many files. The first 
possible solution to this problem is to use special write-once read-many 
devices. Besides the cost of these devices, there is a problem: the 
order of the files in the device is exactly the order in which they are 
written, thus by matching this order is possible to match voter, e.g.\ 
VoteToken, to VoteAuthorization to Vote (obviously if one has read 
access to all devices). To prevent this, we need to add another 
requirement on the file creation:

\begin{itemize}

\item the order in the directory and timestamp of the files should not 
be meaningful.

\end{itemize}

A second solution is to adopt normal disks and to use RSBAC to mimic as 
much as possible the behavior of a write-once read-many filesystem. 
What we can do is to mark as append-only the directories and all 
included files where our files will be written. Moreover the partitions 
where these files reside must be mounted with hard-locking enabled. Thus 
when we write one of these files we open-append it acquiring a hard lock 
only if the file does not exist, and before starting writing we check 
that the file is empty. Doing in this way, the files can be always 
appended later, so it is not a true write-once, but the initial contents 
of the files cannot be modified. We then change the access time and 
modified time of all files in the directory to a fixed timestamp, so 
that all files appear as having been created and modified at the same 
time (this is again possible using the detailed permissions on 
capabilities of RSBAC).

Still we haven't completely solved the problem of the ordering of the 
files in the directories since both the order of the filenames in the 
directory file and the inode numbers leak information about the order of 
creation of the files. In our tests the leaked information appears to be 
little and difficult to use so that at best we have been able to 
reconstruct only a partial order of the files which has not allowed us 
to precisely match files in different directories.

\section{Application setup and security}

The PHP applications require some configuration. The choices give 
different level of security and different features. As always, more 
user-friendly features often correspond to lower security levels.

To configure the AuthSrv we need to add the public and private OpenPGP 
keys already described. The private key of the AuthSysMgr is encrypted 
with a passphrase. The application allows either to write the passphrase 
to a file which is read every time it is needed, or it is loaded by the 
system manager at boot by hand in a opportunely configured gpg-agent. 
The second choice is in principle more secure, since the passphrase is 
not written in a file readable by the apache user, but adds an extra 
failure point. Indeed if the gpg-agent crashes or stops working, the 
full application cannot work either. The same considerations apply to 
the VoteSrv.

It is possible to choose if each username must correspond to a single 
VoteToken, or if username and VoteToken are checked independently. In 
this second case more voters can use the same username, but still each 
voter can use her own VoteToken only once.

It is possible to choose the length of the random string generated by 
the application giving the possibility to balance the trade-off between 
time to generate and space to memorize, likelihood of collisions and so 
on.

Having two independent servers running, a possible problem arises if the 
second server is unreachable, not working or if the voter wants to 
connect to it at a later moment. Since the VoteToken can be used only 
once, how can the voter save her VoteAuthorization for later use?

In the application there are two possibilities on top of the more secure 
but less friendly of not allowing the voter to save her 
VoteAuthorization. (Actually the VoteAuthorization is stored as a hidden 
field in the web Form sent to the voter's browser. By reading the source 
of the page the voter can save her VoteAuthorization. Moreover notice 
that if a client proxy is used, this problem does not arise since the 
client proxy can store securely the VoteAuthorization for the voter.)

The first solution is to offer the possibility to the voter to print (or 
save on disk) her VoteAuthorization. Later on the voter can connect 
directly to the VoteSrv (through the Anonymizer) and type/past in a web 
Form the VoteAuthorization. 

The second possibility is to store the VoteAuthorization on the AuthSrv 
as a file with a name containing the voter's VoteToken. If the voter 
authenticates again to the AuthSrv with her own already used VoteToken, 
no new VoteAuthorization is created but the old VoteAuthorization is 
given to the voter. At the end of the ballot period all stored temporary 
VoteAuthorizations are deleted since they could leak some information 
between the voter (the VoteToken in the filename) and the 
VoteAuthorization (even if these last are encrypted with the VoteSysMgr 
public key).

In both cases, as an added security for the voter, a PIN can be 
generated and added to the VoteAuthorization before encryption. This PIN 
is not stored anywhere else and is only sent to the voter when the 
VoteAuthorization is first created. When the voter later connects to the 
VoteSrv, she will be asked the PIN in the ballot web Form. Thus even if 
someone intercepts or steals the stored VoteAuthorization and tries to 
cast a ballot with it, without the PIN is useless.

The configuration of the VoteSrv is similar to the one of the AuthSrv.

%

%

\section*{Acknowledgments} 

We thank H.\ Bechmann-Pasquinucci for inspiring remarks.

\end{document}